\documentclass[aps,pra,superscriptaddress,twocolumn]{revtex4-1}

\usepackage{amssymb}
\usepackage{graphicx}
\usepackage{amsmath}
\usepackage{physics}
\usepackage{amsthm}
\usepackage{amsfonts}
\usepackage{color}
\usepackage{bm}
\usepackage{bbm}
\usepackage{url}
\usepackage[driverfallback=dvipdfm]{hyperref}
\hypersetup{pdfpagemode=FullScreen,colorlinks=true,breaklinks,urlcolor=blue,linkcolor=blue,citecolor=blue}

\usepackage{txfontsb}
\usepackage[OT1]{fontenc}

\begin{document}

\global\long\def\id{\mathbbm{1}}
\global\long\def\ui{\mathbbm{i}}
\global\long\def\ud{\mathrm{d}}

\title{Preparation and observation of anomalous counterpropagating edge states in a periodically driven optical Raman lattice}

\author{Hongting Hou}
\affiliation{School of Physics and Institute for Quantum Science and Engineering, Huazhong University of Science and Technology, Wuhan 430074, China}
\author{Long Zhang}
\email{lzhangphys@hust.edu.cn}
\affiliation{School of Physics and Institute for Quantum Science and Engineering, Huazhong University of Science and Technology, Wuhan 430074, China}
\affiliation{Hefei National Laboratory, Hefei 230088, China}

\begin{abstract}
Motivated by the recent observation of real-space edge modes with ultracold atoms [Braun {\it et al.}, Nat. Phys. {\bf 20}, 1306 (2024)],
we investigate the preparation and detection of anomalous counterpropagating edge states---a defining feature of the anomalous Floquet valley-Hall (AFVH) phase---in a two-dimensional periodically driven optical Raman lattice. 
Modeling the atomic cloud with a Gaussian wave packet state, we explore, both analytically and numerically, how the population of edge modes depends on the initial-state parameters. In particular, we reveal that, in addition to the internal spin state, the initial momenta parallel and perpendicular to the boundary play essential roles: they independently control the selective population of edge states across distinct momenta and within separate quasienergy gaps. Furthermore, we examine the wave-packet dynamics of counterpropagating edge states and demonstrate that their characteristic motion is robust against long-range disorder. These results establish a theoretical framework for future experimental explorations of the AFVH phase and topological phenomena associated with its unique edge modes.
\end{abstract}

\maketitle

\section{Introduction}

Topological quantum matter has been the focus of intense research over the past decades~\cite{TI_review1,TI_review2,Sato2017_review}. 
Central to the discovery of topological materials~\cite{Konig2007,Hsieh2008,Xia2009,Chang2013,Xu2015,Lv2015} is the bulk–boundary correspondence~\cite{Hatsugai1993,Qi2006,Mong2011,Essin2011}, 
which connects nontrivial bulk topology to the existence of gapless boundary modes. 
Recently, this foundational principle has been modified and extended to periodically driven systems~\cite{Rudner2013,Nathan2015}. 
It has been shown that due to the periodicity of quasienergy, traditional topological invariants of Floquet bands, 
such as the Chern numbers in two-dimensional systems, are insufficient for fully classifying the Floquet topology. 
Instead, information about edge modes can be derived from the winding numbers defined in the higher momentum-time space, 
which characterize the topology of the quasienergy gaps~\cite{Rudner2013}. 
This new formulation of bulk-boundary correspondence reveals that periodically driven systems can host anomalous edge states 
and exhibit novel Floquet topological phases that lack static counterparts~\cite{Kitagawa2010,Carpentier2015,Fruchar2016,Roy2017,Yao2017,Morimoto2017,Huang2023,Ghosh2024}.

With their exceptional controllability and cleanliness, ultracold atoms have emerged as an ideal platform for simulating a variety of topological models~\cite{Goldman2016_review,Zhang2018_review,Cooper2019_review,Schafer2020_review}, including the one-dimensional (1D) Su-Schrieffer-Heeger chain~\cite{Atala2013}, chiral topological phases~\cite{Song2018}, two-dimensional (2D) Chern insulators~\cite{Aidelsburger2013,Miyake2013,Jotzu2014,Aidelsburger2015,Wu2016,Sun2018a,Liang2023}, and three-dimensional topological semimetals~\cite{Song2019,Wang2021}. These advancements lay a crucial foundation for exploring Floquet topological phases in periodically driven optical lattices~\cite{Eckardt2017_review,Rudner2020_review,Weitenberg2021_review}. Recently, several experimental groups have realized anomalous Floquet topological bands~\cite{Wintersperger2020,Lu2022,Zhang2023}. Notably, Zhang  {\it et al.}~\cite{Zhang2023} successfully manipulated the Floquet band topology by adjusting driving-induced band crossings, unveiling a nontrivial phase known as the anomalous Floquet valley-Hall (AFVH) phase. This phase exhibits protected counterpropagating edge states within each quasienergy gap, eluding classification by conventional global topological invariants~\cite{Zhang2022,Lababidi2014,Umer2020}. In their experiment, Zhang {\it et al.} identified this unconventional phase using a quench protocol based on local Floquet band structures~\cite{Zhang2023,Zhang2020}; however, its hallmark feature---the counterpropagating edge states---has yet to be observed, largely due to the challenges associated with creating well-defined boundaries in ultracold atomic gases, necessitating careful preparation and probing techniques for edge modes~\cite{Stanescu2009,Stanescu2010,Buchhold2012,Goldman2012,Goldman2013,Reichl2014,Goldman2016,Leder2016,Martinez2023,Wang2024}.

Recently, Braun {\it et al.}~\cite{Braun2024} achieved an experimental breakthrough in realizing and observing topological edge modes in real-space Floquet systems.
Compared to previous observations in synthetic dimensions~\cite{Mancini2015,Stuhl2015,Chalopin2020}, Braun {\it et al.}~\cite{Braun2024} generated and controlled a steep system edge using a programmable repulsive optical potential, 
and loaded atoms into real-space edge modes by positioning the atomic cloud directly at the edge with an optical tweezer.
Upon releasing the atoms, they observed unidirectional wave-packet propagation along the potential boundary,
providing clear confirmation of the population of chiral edge modes~\cite{Martinez2023,Braun2024}.
This experimental study opens avenues for exploring edge-state dynamics and topological phenomena in real-space platforms.

Inspired by recent experimental advances~\cite{Zhang2023,Braun2024}, we present in this work 
an analytical and numerical investigation into the preparation and observation of counterpropagating edge states in periodically driven ultracold atoms, 
providing a direct and explicit approach to identifying the AFVH phase. 
We analyze how edge-state preparation is influenced by initial-state parameters and find that, in addition to the internal spin state, 
the momentum of the wave packet plays a crucial role.
Specifically, when populating edge modes in the AFVH phase,
the initial momentum parallel to the boundary determines which edge state is effectively populated at different momenta, 
while the initial momentum perpendicular to the boundary serves as a control parameter for selecting
edge states within one of the two quasienergy gaps.
By preparing an initial state that simultaneously populates two oppositely chiral edge states, 
we numerically investigate the resulting wave-packet dynamics, which exhibits a counterpropagating motion characteristic of these edge states.
Additionally, we demonstrate that the counterpropagating edge-state transport in the AFVH phase remains robust in the presence of long-range disorder.
The preparation and detection scheme outlined in this work is well-suited for implementation with current experimental techniques.

This paper is organized as follows. 
In Sec.~\ref{Sec2}, we introduce the Floquet topological model realized in a 2D shaken optical Raman lattice and analyze its anomalous 
topological phases.
In Sec.~\ref{Sec3}, we demonstrate the method of populating a target edge mode by illustrating how the initial-state parameters, particularly the initial momenta parallel and perpendicular to the boundary, affect the edge-state preparation.
In Sec.~\ref{Sec4}, we investigate the wave-packet transport of counterpropagating edge modes in the AFVH phase and demonstrate their robustness against long-range disorder. 
A brief discussion and summary are presented in Sec.~\ref{Sec5}.
More details are given in the Appendixes.

\begin{figure*}
\includegraphics[width=0.99\textwidth]{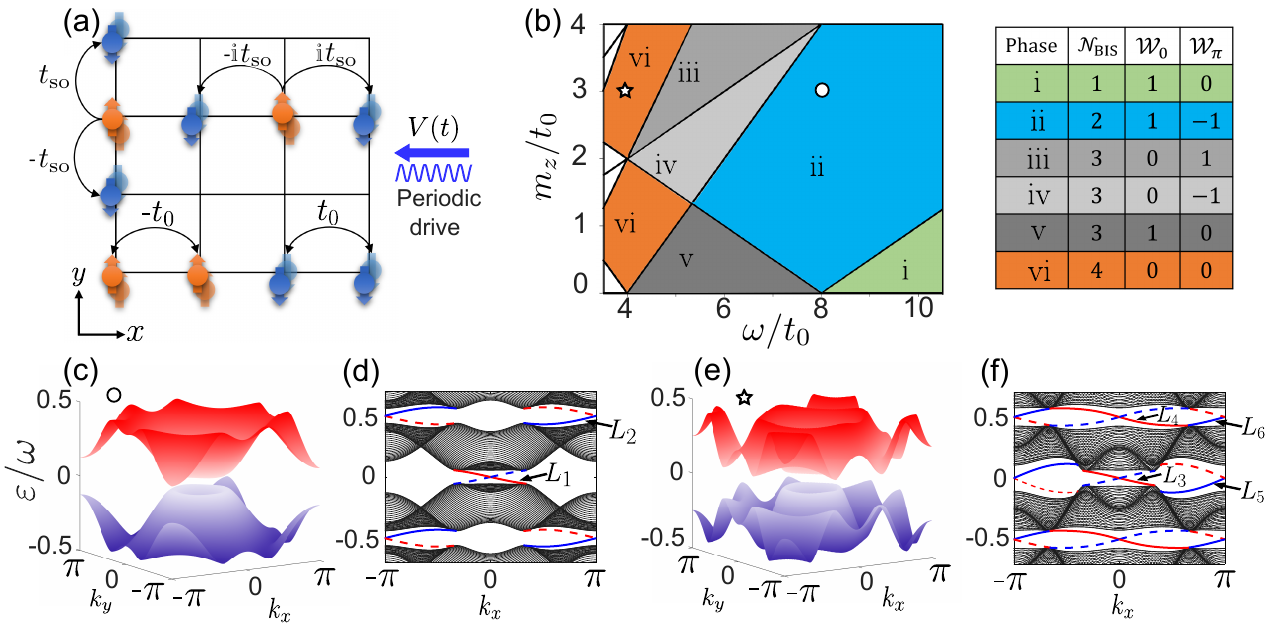}
    \caption{(a) Sketch diagram of the 2D Floquet topological model realized in a shaken optical Raman lattice. (b) Phase diagram as function of the driving frequency $\omega$ and the Zeeman constant $m_z$. The Floquet phases can be distinguished by two topological invariants ${\cal W}_{0,\pi}$ combined with the number of BISs ${\cal N}_{\rm BIS}$. 
Here, we set $t_{\rm so}=0.5t_0$ and  $V_{0}=3t_0$.
(c)-(f) Quasienergy band structures with periodic boundary conditions [(c) and (e)] or for a cylindrical geometry [(d) and (f)], corresponding to two phases in {\color{red} (b)} labeled by circle ($m_z=3t_0$, $\omega=8t_0$) and star ($m_z=3t_0$, $\omega=4t_0$) markers, respectively. 
In the open-boundary spectrums [(d) and (f)], edge modes at the left (right) boundary are plotted as solid (dashed) curves, and colored in red (blue) for negative (positive) chirality.
}\label{fig1}
\end{figure*}

\section{Model}\label{Sec2}

We consider the Floquet topological system that has been recently realized with ultracold bosons trapped in a shaken optical Raman lattice~\cite{Zhang2023}, as depicted in Fig.~\ref{fig1}(a).
The optical Raman lattice scheme has been employed to realize many topological models~\cite{Song2018,Wu2016,Sun2018a,Liang2023,Song2019,Wang2021,Zhang2023,LiuXJ2013,LiuXJ2014,LiuXJ2016,ZhangDW2016,Lang2017,Zhang2017,WangBZ2018,Yang2018,LiuH2019,Zheng2019,Wang2020,Hou2020,LuWang2020,WangBB2021,WangY2022,Ziegler2022,LiH2023,Zhou2023,Zhao2023,ZhangH2024,WangJT2024}.
The time-periodic Hamiltonian reads
\begin{align}\label{Ham_real}
 H(t)=&-t_{\rm 0}\sum_{\langle {\bf r},{\bf r}' \rangle }\left(c^\dag_{{\bf r}\uparrow}c_{{\bf r}'\uparrow}-c^{\dagger}_{{\bf r}\downarrow}c_{{\bf r}'\downarrow}\right)\nonumber\\
  &+\sum_{{\bf r}}\left[m_{z}+V(t)\right]\left(n_{{\bf r}\uparrow}-n_{{\bf r}\downarrow}\right)\nonumber\\
    &+\sum_{x}\left[\ui t_{\rm{so}}\left(c_{x+1\downarrow}^{\dagger}c_{x\uparrow}-c_{x-1\downarrow}^{\dagger}c_{x\uparrow}\right)+\rm{h.c.}\right]\nonumber \\
    &+\sum_{y}\left[t_{\rm{so}}\left(c_{y+1\downarrow}^{\dagger}c_{y\uparrow}-c_{y-1\downarrow}^{\dagger}c_{y\uparrow}\right)+\rm{h.c.}\right],
\end{align}
where $c_{{\bf r}\sigma}$ ($c^\dag_{{\bf r}\sigma}$) are the annihilation (creation) operators for spin $\sigma=\uparrow,\downarrow$ at lattice site ${\bf r}=(x,y)$, 
$n_{{\bf r}\sigma}\equiv c^{\dagger}_{{\bf r}\sigma}c_{{\bf r}\sigma}$,
$t_0$ ($t_{\rm so}$) denotes the spin-conserved (-flipped) hopping amplitude along the $x$- or $y$-direction, and the periodic drive $V(t)=2V_0\cos\omega t$ is applied to the Zeeman term, with $\omega$ being the driving frequency. 
For simplicity, we set the lattice constant as the unit of length, namely, $a=1$ and assume that the parameters $t_0$, $t_{\rm so}$, and $m_z$ are all positive.
The Bloch Hamiltonian in momentum space takes the form
\begin{align}~\label{FHam2D}
\begin{split}
H({\bf k}, t)=H_{\rm s}({\bf k})+V(t)\sigma_z,\quad H_{\rm s}({\bf k})={\bf h}({\bf k})\cdot{\bm\sigma},
\end{split}
\end{align}
where $\sigma_{x,y,z}$ are the Pauli matrices, and ${\bf h}({\bf k})=(h_x,h_y,h_z)=(2t_{\rm so}\sin k_x, 2t_{\rm so}\sin k_y,m_z-2t_0\cos k_x-2t_0\cos k_y)$
describes a quantum anomalous Hall model~\cite{Wu2016,Sun2018a}. 
This periodically driven system hosts rich Floquet topological phases beyond conventional classification. The phase diagram as function of $\omega$ and $m_z$ is depicted in Fig.~\ref{fig1}(b).
In the following, we shall review how to characterize the Floquet topology and subsequently apply the approach to analyze anomalous Floquet topological phases.

A Floquet system can be described by an effective Hamiltonian $H_{F}=\ui\ln{U(T)}/T$, whose eigenvalues $\varepsilon({\bf k})$ form the Floquet bands with two inequivalent quasienergy gaps~\cite{Rudner2020_review}. Here $T=2\pi/\omega$ is the drive period and $U(T)$ is the time-evolution operator over one period, defined by
\begin{equation}\label{UT_def}
U(T)={\cal T}e^{-\ui\int_{0}^{T}H(t)dt},
\end{equation}
where ${\cal T}$ denotes the time ordering.
For a driven system described by Eq.~\eqref{FHam2D}, the Floquet Hamiltonian also take a Dirac-type form 
\begin{align}\label{HF_general}
H_{F}(\bold{k})=\sum_{\alpha=x,y,z}h_{F, \alpha}(\bold{k})\sigma_\alpha.
\end{align}
We adopt the concept of band-inversion surfaces (BISs)
and employ the generalized bulk-surface duality to characterize the Floquet topology (see Appendix~\ref{App1} and Ref.~\cite{Zhang2020} for details).
Here, for a 2D Floquet system, a BIS refers to a 1D closed curve in the first Brillouin zone where band inversion occurs, i.e.,
\begin{align}~\label{FBIS_def}
{\rm BIS}\equiv\{\bold{k}| h_{F, z}(\bold{k})=0\}.
\end{align}
Unlike static systems, Floquet band crossings can appear at both $\varepsilon=0$ and $\varepsilon=\pi/T$. 
We name the gap around the quasienergy $0$ ($\pi /T$) as the $0$ gap ($\pi$ gap) and the BIS  associated with this gap as $0$ BIS ($\pi$ BIS).
Particularly, for the Hamiltonian \eqref{FHam2D}, the definition \eqref{FBIS_def} reduces to $h_z(k)=n\omega/2$, 
where $n=0,1,2,\cdots$, and a 0 BIS ($\pi$ BIS) corresponds to $n$ being even (odd).
Hereafter, we will refer to a BIS corresponding to a nonnegative $n$ as the one of {\it order} $n$.
The Floquet band topology is contributed by all $0$ and $\pi$ BISs~\cite{Zhang2020}:
\begin{equation}~\label{W_Floquet}
\mathcal{C}=\mathcal{W}_0-\mathcal{W}_{\pi}, \quad \mathcal{W}_q = \sum_{i}\nu_{i}^{(q)}.
\end{equation}
Here $\nu_{i}^{(q)}$ represents the topological invariant defined on the $i$th $q$ BIS ($q = 0, \pi$), $\mathcal{C}$ is the Chern number of the Floquet bands, 
and ${\cal W}_{0}$ (${\cal W}_{\pi}$) characterizes the number of boundary modes inside the 0 gap ($\pi$ gap).


Based on the classification theory presented above, 
three basic features about the 2D periodically driven cold-atom system have been demonstrated both theoretically~\cite{Zhang2022} and experimentally~\cite{Zhang2023}:
(i) The two types of BISs---0 BIS and $\pi$ BIS---always appear alternatively.
(ii) The topological invariant $\nu_{i}^{(q)}$ can be uniquely determined by the BIS configuration in the first Brillouin zone. 
Specifically, $\nu_{i}^{(q)}=+1$ ($-1$) when the BIS surrounds the $\Gamma$ ($M$) point.
(iii) Each $q$ BIS with $\nu_{i}^{(q)}\neq0$ ($q=0,\pi$) corresponds to a gapless chiral edge mode within the $q$ gap, 
rendering the so-called BIS-boundary correspondence~\cite{Zhang2022}.
This correspondence is reflected in the effective Hamiltonian $H_{\rm eff}^{(n)}({\bf k})=\widetilde{\bf h}({\bf k})\cdot{\bm\sigma}$, which is introduced to describe the emergence of a driving-induced BIS of order $n$ (see Appendixes~\ref{App1} and \ref{App2}), 
where $\widetilde{\bf h}({\bf k})=(2\tilde{t}_{\rm so}\sin k_x, 2\tilde{t}_{\rm so}\sin k_y,\widetilde{m}_z-2t_0\cos k_x-2t_0\cos k_y)$, with
\begin{align}\label{mz_tso_eff}
\widetilde{m}_z=m_z-\frac{n\omega}{2},\quad \tilde{t}_{\rm so}=(-1)^nJ_n\left(\frac{4V_0}{\omega}\right)t_{\rm so}.
\end{align}
Here $J_n(z)$ denotes the Bessel function of the first kind of order $n$.
These three features enable complete characterization of the Floquet band topology via BIS configurations.
Equation~\eqref{mz_tso_eff} demonstrates that each BIS contributes topological features through an effective Hamiltonian
structurally replicating $H_{\rm s}({\bf k})$ [cf. Eq.~\eqref{FHam2D}] under parameter substitutions.
The original $H_{\rm s}({\bf k})$ exhibits two topological phases: a Chern number ${\cal C}=+1$ phase for $0<m_z<4t_0$
and a ${\cal C}=-1$ phase for $-4t_0<m_z<0$. Consequently, the Floquet phase boundaries are governed by  $|m_z-n\omega/2|=4t_0$, corresponding to the creation or annihilation of a BIS, and $|m_z-n\omega/2|=0$, marking the switching of BIS geometry between $\Gamma$-centered and $M$-centered configurations.
Furthermore, we can identify two novel Floquet topological phases, as detailed below.

The phase labeled “ii”, characterized by the number of BISs ${\cal N}_{\rm BIS}=2$, is an anomalous phase with a high Chern number ${\cal C}=2$.
As illustrated in Fig.~\ref{fig1}(c), the quasienergy bands in phase ii exhibit two ring-shaped structures where the band inversion occurs, resulting in two BISs. One is a $0$ BIS circling the $\Gamma$ point, and the other is a $\pi$ BIS circling the $M$ point. According to the feature (ii), we have ${\cal W}_0=\nu_{1}^{(0)}=+1$, ${\cal W}_\pi=\nu_{1}^{(\pi)}=-1$, and thus ${\cal C}={\cal W}_0-{\cal W}_\pi=2$.
The feature (iii) predicts that there exists a chiral edge mode in either quasienergy gap, which is confirmed by the open-boundary spectrum shown in Fig.~\ref{fig1}(d).

The phase labeled “vi” with ${\cal N}_{\rm BIS}=4$ is the AFVH phase, which evades the conventional classification by global topological invariants ${\cal W}_{0,\pi}$~\cite{Zhang2022}.  
The quasienergy band structure shown in Fig.~\ref{fig1}(e) indicates the presence of four BISs, of which two are $0$ BISs and two are $\pi$ BISs.
Furthermore, the inner $0$ BIS ($\pi$ BIS) circles the $\Gamma$ point, and the outer $0$ BIS ($\pi$ BIS) circles the $M$ point.
Based on the BIS circling configuration, we have $\nu_{1}^{(0)}=\nu_{1}^{(\pi)}=+1$ and $\nu_{2}^{(0)}=\nu_{2}^{(\pi)}=-1$, which yields ${\cal W}_0=\nu_{1}^{(0)}+\nu_{2}^{(0)}=0$,
${\cal W}_\pi=\nu_{1}^{(\pi)}+\nu_{2}^{(\pi)}=0$, and ${\cal C}={\cal W}_0-{\cal W}_\pi=0$.
Despite the trivial bulk topological invariants, 
the BIS-boundary correspondence ensures the presence of counterpropagating edge modes in both quasienergy gaps, as illustrated in Fig.~\ref{fig1}(f).
Although this unconventional phase has been identified through quench dynamics~\cite{Zhang2023}, its unique edge modes have yet to be experimentally observed.
In the following sections, we delve into the scheme for preparing and detecting the counterpropagating edge states.

\section{Edge-state preparation}\label{Sec3}

In this section, we develop and numerically validate a protocol for selectively populating edge modes residing in different quasienergy gaps.

\subsection{Gaussian wave packet and the internal state}

\begin{figure}
    \centering
    \includegraphics[width=0.45\textwidth]{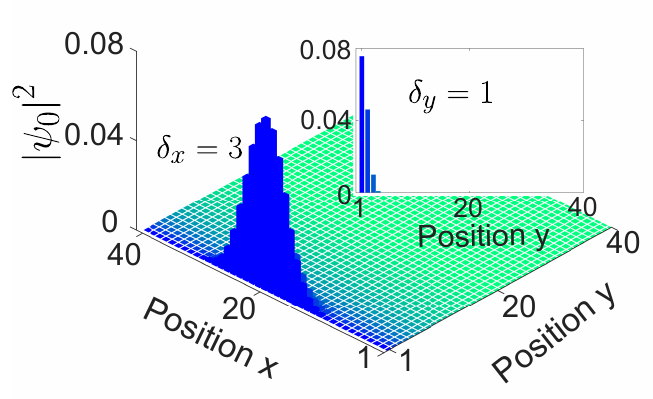}
    \caption{A typical example of the density distribution of the initial state prepared in a $40\times40$ square lattice with the parameters $(x_0,y_0)=(20,1)$ and $(\delta_x,\delta_y)=(3,1)$. In our settings, periodic (open) boundary conditions are imposed in the $x$ ($y$)-direction. }
    \label{fig2}
\end{figure}

We consider an initial state with a Gaussian spatial distribution centered at $(x_0,y_0)$ of width $\delta_{x,y}$:
\begin{align}\label{inista}
    &\ket{\Psi_0}=\psi_0(x,y)\ket{\eta}\nonumber\\
    &=\frac{1}{\mathcal{N}}\exp[\frac{-(x-x_0)^2}{4\delta_x^2}+\ui q_xx+\frac{-(y-y_0)^2}{4\delta_y^2}+\ui q_yy]\ket{\eta}
\end{align}
where $\mathcal{N}$ denotes the normalization factor and $\ket{\eta}$ is a two-component spinor. 
The initial momentum $(q_x , q_y)$ serves as a control parameter for edge-state preparation, as will be elaborated below.
For simplicity, we adopt a cylindrical geometry with periodic boundary conditions in the $x$-direction and open boundary conditions in the $y$-direction, 
and fix $x_0=L/2$ and $y_0=1$ throughout this study ($L$ is the lattice size).
An example of the wave packet $|\psi_0(x,y)|^2$ is shown in Fig.~\ref{fig2}.
While our discussion focuses on the preparation of edge modes at the left boundary, the results can readily be extended to other cases.

We calculate the overlaps $P_n=\sum_{x,y}\abs{\braket{\phi_n}{\Psi_0}}^2$ and investigate the real-space chiral motion of the wave packet to examine the edge-state preparation, where $\ket{\phi_n}$ denote the eigenstates of the Floquet Hamiltonian.
The time-evolving state at stroboscopic time $t=nT$ ($n\in\mathbb{Z}$) is given by
\begin{equation}
    \ket{\Psi(t=nT)}=e^{-\ui H_F\cdot nT}\ket{\Psi_0}=
      \begin{pmatrix}
       \psi_\uparrow(x,y,t) \\   
       \psi_\downarrow(x,y,t)
       \end{pmatrix}.
\end{equation}
The wave-packet propagation along the edge can be observed by calculating the distribution at the boundary, i.e.,
\begin{equation}
    \abs{\Psi(x,t)}^2\overset{\rm def}{=}\sum_{\sigma=\uparrow,\downarrow}~\sum_{y_{\rm edge}}~\abs{\psi_{\sigma}(x,y,t)}^2,
\end{equation}
where $y_{\rm edge}$ denotes the layers at the left boundary, e.g. $y_{\rm edge}=1,2,3,4$ in our calculations.

To effectively populate an edge mode, both the internal spin state and the spatial wave packet should be carefully prepared.
 Analytical results show that the left edge states take the form
\begin{align}\label{phiL}
\ket{\phi_L(k_x,y)}=\frac{u_{k_x}(x)}{\mathcal{N}_L}\left(\lambda_{L+}^{y}-\lambda_{L-}^{y}\right)\ket{\eta_L},
\end{align}
where $\mathcal{N}_L$ denotes the normalization factor, $u_{k_x}(x)$ is the Bloch wave function along the $x$-direction, and
the complex variables $\lambda_{L\pm}$ are given in Appendix~\ref{App2}.
The internal state $\ket{\eta_L}=\frac{1}{\sqrt{2}}(1,-1)^{\sf T}$ or  $\frac{1}{\sqrt{2}}(1,1)^{\sf T}$, depending on the system parameters.
Particularly, when $V_0/\omega<0.6$, we have a general conclusion (see Appendix~\ref{App2} for details)
\begin{align}\label{etaL}
\ket{\eta_L}
=\left\{
\begin{array}{ll}
\frac{1}{\sqrt{2}}(1,-1)^{\sf T}  & \textrm{0-gap edge modes}\\
\frac{1}{\sqrt{2}}(1,1)^{\sf T}  & \textrm{$\pi$-gap edge modes}
\end{array}\right. .
\end{align}
However, it should be noted that the AFVH phase under consideration, marked by a star in Fig.~\ref{fig1}(b), is characterized by $V_0=3t_0$ and $\omega=4t_0$, which does not satisfy $V_0/\omega<0.6$. Consequently, the left edge mode labeled ``$L_3$'' resides in the 0 gap and possesses the internal state $\ket{\eta_L}=\frac{1}{\sqrt{2}}(1,1)^{\sf T}$; other left edge modes are in agreement with the prediction given in Eq.~\eqref{etaL} (see Appendix~\ref{App2}).
The expression \eqref{phiL} indicates that, to populate a desired edge mode, the atoms must be initialized in the appropriate internal state $\ket{\eta}=\ket{\eta_L}$. In the following, we assume that this internal state preparation has been successfully achieved via a carefully applied Raman-coupling pulse, 
which enables control over the population of edge states by finely tuning the wave packet's momentum $(q_x,q_y)$.

\subsection{Tuning the initial momentum $(q_x,q_y)$}

An edge state is characterized by both the momentum $k_x$ and the quasienergy gap it resides. We will later demonstrate that the initial momentum $q_x$ ($q_y$), which is parallel (perpendicular) to the boundary, serves as a control parameter for populating edge modes at different momenta (within distinct gaps). 
We will examine both the ${\cal C}=2$ phase and the AFVH phase to illustrate it.

\begin{figure}
    \includegraphics[width=0.50\textwidth]{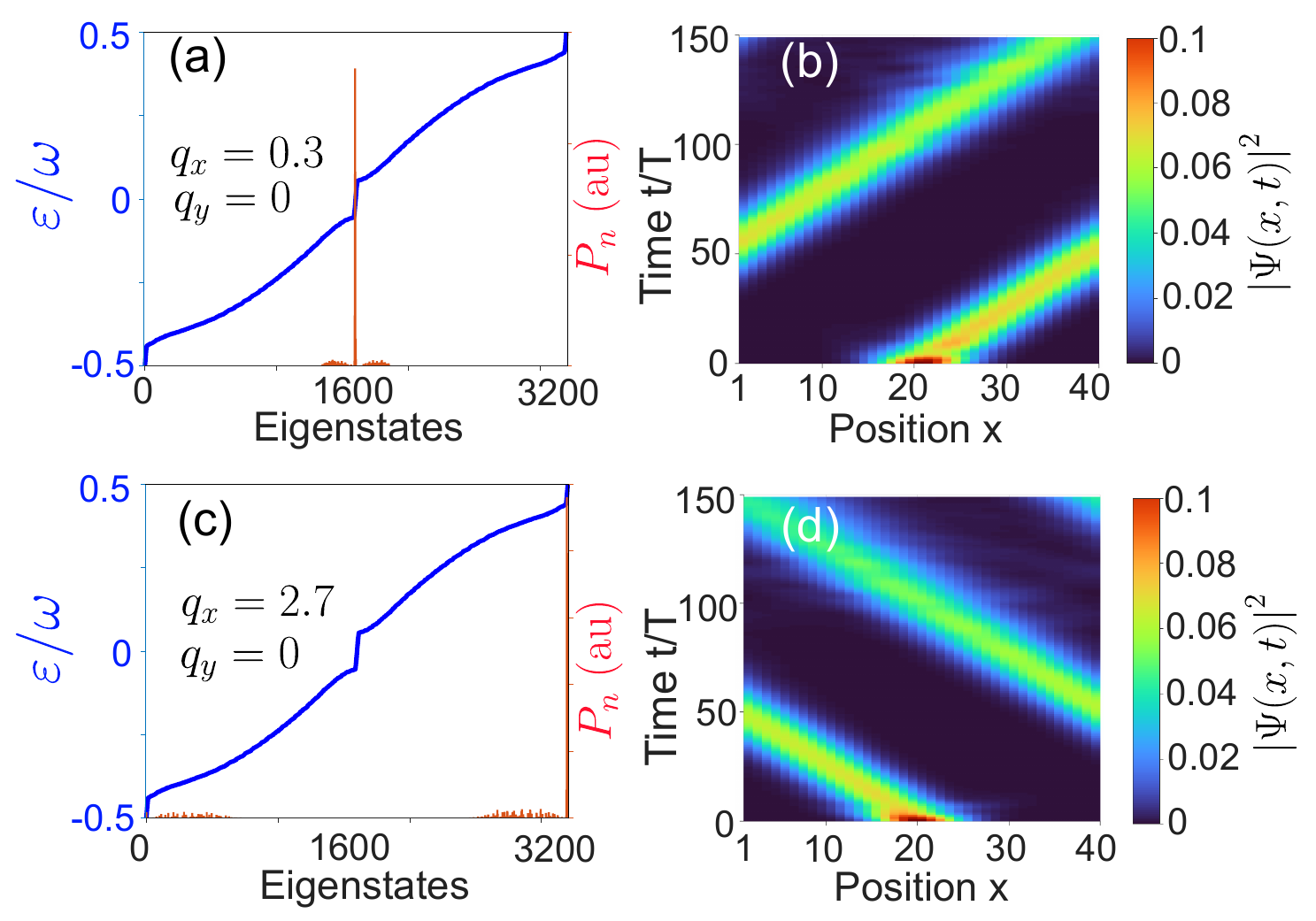}
    \caption{Edge-state preparation in the ${\cal C}=2$ phase, with the two left edge modes $L_{1,2}$, as labeled in Fig.~\ref{fig1}(d), being considered. Overlaps of the initial state with Floquet eigenstates $P_n$ are shown in (a) and (c), while (b) and (d) illustrate how the initial-state wave packet propagates along the boundary. 
  In (a) and (b), the initial state is prepared in the spin state $\ket{\eta}=\frac{1}{\sqrt{2}}(1,-1)^{\sf T}$ with an initial momentum $(q_x,q_y)=(0.3,0)$.
  In (c) and (d), the initial state is prepared in $\ket{\eta}=\frac{1}{\sqrt{2}}(1,1)^{\sf T}$ with $(q_x,q_y)=(2.7,0)$.
  Here, we take $t_{\rm so}=0.5t_0$, $m_{z}=3t_0$, $V_0=3t_0$, $\omega=8t_0$, and $(\delta_x,\delta_y)=(3,1)$.}
    \label{fig3}
\end{figure}

As illustrated in Fig.~\ref{fig1}(d), the two left edge modes in the ${\cal C}=2$ phase are well separated in momentum $k_x$:
The edge mode in the 0 gap, labeled ``$L_1$'', is situated around $k_x=0$, 
while the edge mode in the $\pi$ gap, labeled ``$L_2$'', emerges around $k_x=\pi$.
This configuration suggests that an initial kick with varying $q_x$ can be employed to control the edge-state preparation. 
In Fig.~\ref{fig3}, we initialize two states that are localized at the center of the edge along the $y$-direction, corresponding to momenta 
$(q_x,q_y)=(0.3,0)$ and $(q_x,q_y)=(2.7,0)$, respectively. 
The overlaps presented in Figs.~\ref{fig3}(a) and \ref{fig3}(c) reveal that these initial states are aimed at the edge modes within distinct gaps. 
Meanwhile, Figs.~\ref{fig3}(b) and \ref{fig3}(d) depict the chiral motion of the wave packets, confirming that the edge modes with opposite chirality are effectively populated.


In the AFVH phase, however, two left edge modes residing in different gaps can emerge in the same momentum region.
As shown in Fig.~\ref{fig1}(f), the edge modes labeled ``$L_3$'' and ``$L_4$''  (``$L_5$'' and ``$L_6$'' ) are both located around $k_x=0$ ($k_x=\pi$), 
and cannot be selectively populated by adjusting the initial momentum $q_x$. 
Surprisingly, we find that the initial momentum perpendicular to the boundary, $q_y$, can be employed to effectively control the population of edge states within a specific gap.
This finding can be clarified by analyzing the spatial distribution of the edge state phase $\theta(y)$ along the y-direction.

\subsubsection{Analytical study}

We first consider left edge states at $k_x=0$, with the wave-packet function along the $y$-direction taking the form
\begin{align}~\label{phi_y1}
\phi_L(y)=\frac{1}{\mathcal{N}_L}\left(\lambda_{L+}^{y}-\lambda_{L-}^{y}\right),
\end{align}
where, according to Appendix~\ref{App2},
\begin{align}
\lambda_{L\pm}=\frac{\widetilde{m}_z-2t_0\pm\sqrt{\widetilde{m}_z^2-4t_0\widetilde{m}_z+4\tilde{t}_{\rm so}^{\,2}}}{2(t_0+|\tilde{t}_{\rm so}|)},
\end{align}
and $0<\widetilde{m}_z<4t_0$. Here, $\widetilde{m}_z$ and $\tilde{t}_{\rm so}$ are the parameters of the effective Hamiltonian $H_{\rm eff}^{(n)}({\bf k})$, 
as presented in Eq.~\eqref{mz_tso_eff}.
When $\widetilde{m}_z^2-4t_0\widetilde{m}_z+4\tilde{t}_{\rm so}^{\,2}<0$, we write $\lambda_{L,\pm}=|\lambda_{L}|e^{\pm\ui\theta_{L}}$, where 
\begin{align}
|\lambda_{L}|=\sqrt{\frac{t_0-|\tilde{t}_{\rm so}|}{t_0+|\tilde{t}_{\rm so}|}}
\end{align}
and
\begin{align}
\theta_L=\left\{
\begin{array}{ll}
\theta_L^{(0)} & 2t_0\leq \widetilde{m}_z<4t_0\\
\pi-\theta_L^{(0)} & 0<\widetilde{m}_z<2t_0
\end{array}\right. ,
\end{align}
with $\theta_L^{(0)}\equiv\arcsin\left[\sqrt{1-(\widetilde{m}_z-2t_0)^2/(4t_0^2-4\tilde{t}_{\rm so}^2)}\right]$.
With this result, we have
\begin{align}~\label{phi_y2}
\phi_L(y)=\frac{2}{\mathcal{N}_L}|\lambda_{L}|^y\left|\sin(\theta_Ly)\right|e^{\ui\theta(y)},
\end{align}
with the $y$-dependent phase given by
\begin{align}
\theta(y)&=\frac{\pi}{2}\,{\rm sgn}\left[\sin(\theta_Ly)\right] \nonumber\\
&=\left\{
\begin{array}{ll}
\frac{\pi}{2}\,{\rm sgn}\left[\sin(\theta^{(0)}_Ly)\right]  & 2t_0\leq \widetilde{m}_z<4t_0\\
\frac{\pi}{2}\,(-1)^{y+1}{\rm sgn}\left[\sin(\theta^{(0)}_Ly)\right] & 0<\widetilde{m}_z<2t_0
\end{array}\right. .
\end{align}
Here, we focus on the two leftmost sites, i.e., $y=1,2$, the spatial distribution at which dominantly determines the overlaps of the initial state with the edge states. 
We see that when $\theta^{(0)}_L<\pi/2$, the phase distribution $\theta(y)$ satisfies 
\begin{align}
\theta(2)
=\left\{
\begin{array}{ll}
\theta(1)  & 2t_0<\widetilde{m}_z<4t_0\\
-\theta(1) & 0<\widetilde{m}_z<2t_0
\end{array}\right. ,
\end{align}
which means that when $\widetilde{m}_z>2t_0$ ($\widetilde{m}_z<2t_0$), the wave-packet function $\phi_L(y)$ is symmetrically (antisymmetrically) distributed at the two leftmost sites.
Consequently, while an edge state with $\widetilde{m}_z>2t_0$ can be directly populated without an initial momentum $q_y$, 
a kick with $q_y=\pi$ is required to effectively populate an edge state with $\widetilde{m}_z<2t_0$, which is equivalent to introducing a phase factor $e^{i\pi y}$ to the wave packet. 

For left edge states at $k_x=\pi$, we can draw similar conclusions. The wave-packet function $\phi_L(y)$ of an edge state also takes the form of Eq.~\eqref{phi_y1}, but with
\begin{align}
\lambda_{L\pm}=\frac{\widetilde{m}_z+2t_0\pm\sqrt{\widetilde{m}_z^2+4t_0\widetilde{m}_z+4\tilde{t}_{\rm so}^{\,2}}}{2(t_0+|\tilde{t}_{\rm so}|)},
\end{align}
and $-4t_0<\widetilde{m}_z<0$.
When $\widetilde{m}_z^2+4t_0\widetilde{m}_z+4\tilde{t}_{\rm so}^{\,2}<0$, one can write $\phi_L(y)$ in the form of Eq.~\eqref{phi_y2}, 
with the $y$-dependent angle now given by
\begin{align}\label{thetay_pi}
\theta(y)
=\left\{
\begin{array}{ll}
\frac{\pi}{2}\,{\rm sgn}\left[\sin(\theta^{(\pi)}_Ly)\right]  & -2t_0\leq \widetilde{m}_z<0\\
\frac{\pi}{2}\,(-1)^{y+1}{\rm sgn}\left[\sin(\theta^{(\pi)}_Ly)\right] & -4t_0<\widetilde{m}_z<-2t_0
\end{array}\right.,
\end{align}
where $\theta_L^{(\pi)}\equiv\arcsin\left[\sqrt{1-(\widetilde{m}_z+2t_0)^2/(4t_0^2-4\tilde{t}_{\rm so}^2)}\right]$.
 A similar deduction can be then employed to infer that at $k_x=\pi$, an initial kick with $q_y=\pi$ is required to populate an edge state with $\widetilde{m}_z<-2t_0$.

For the AFVH phase, the appearance of four counterpropagating edge modes $L_{3,4,5,6}$, as observed in Fig.~\ref{fig1}(f), requires the corresponding Zeeman constants $\widetilde{m}_{z,i}$ ($i=3,4,5,6$) fall into four different ranges, 
i.e., $4t_0>\widetilde{m}_{z,3}>2t_0>\widetilde{m}_{z,4}>0>\widetilde{m}_{z,5}>-2t_0>\widetilde{m}_{z,6}>-4t_0$.
This, combined with the above results, shows that the initial momentum $q_y$ 
serves as a control parameter for populating the edge modes of the AFVH phase within different gaps:
While the 0-gap edge modes $L_{3,5}$ can be directly populated, an initial kick with $q_y=\pi$ is required to 
effectively populate the edge modes $L_{4,6}$ within the $\pi$ gap.

 \subsubsection{Numerical demonstration}
 
 \begin{figure*}
    \includegraphics[width=0.99\textwidth]{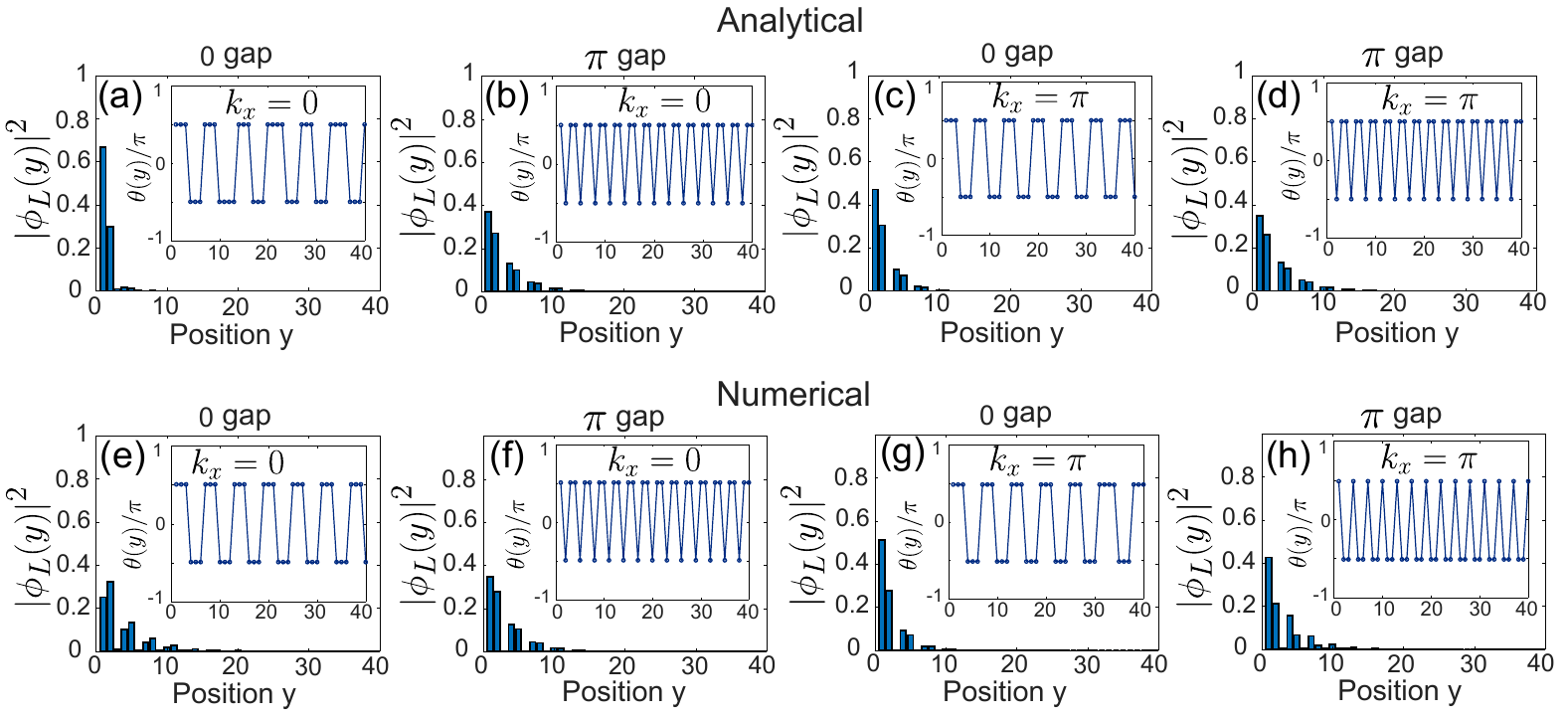}
    \caption{Density distributions $|\phi_L(y)|^2$ of the left edge states in the AFVH phase, located at $k_x=0$ or $\pi$ within the $0$ or $\pi$ gap.  
    For each edge state, the analytical solution (top) is juxtaposed with the corresponding numerical solution (bottom) for a comparison. The inserts show the corresponding phase distributions $\theta(y)$ in the $y$-direction.
     Here, we take $t_{\rm so}=0.5t_0$, $m_{z}=3t_0$, $V_0=3t_0$, and $\omega=4t_0$.}
    \label{fig4}
\end{figure*}

 \begin{figure*}
    \includegraphics[width=\textwidth]{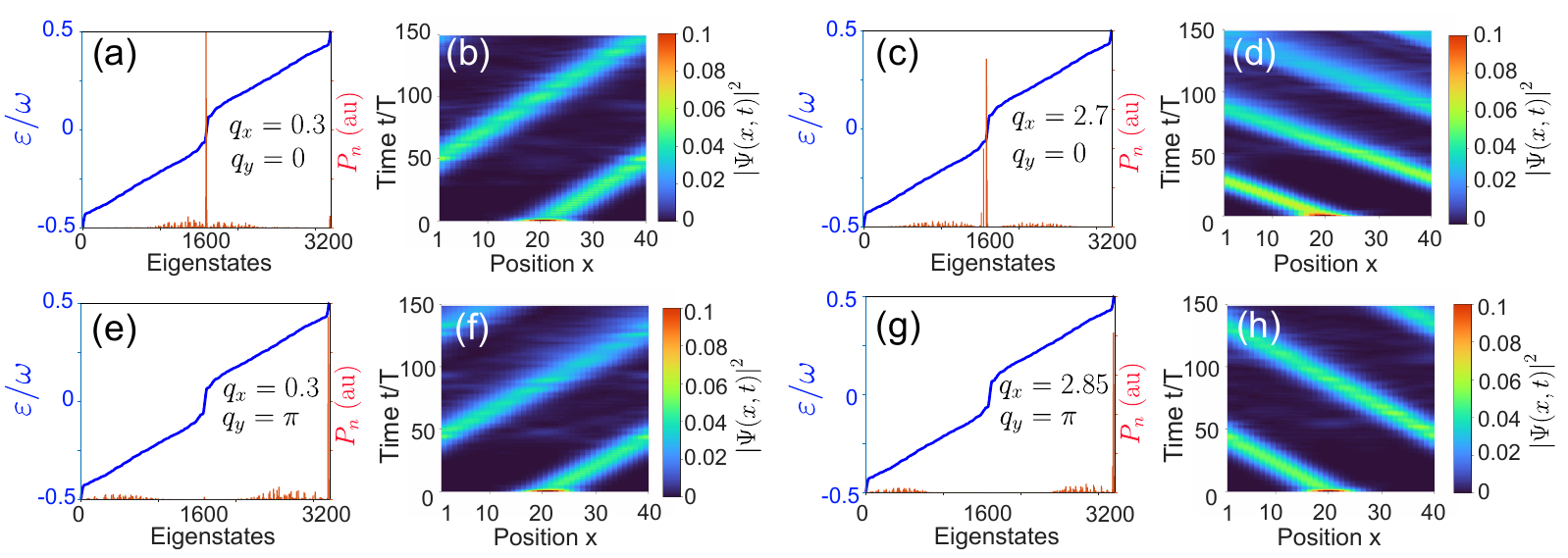}
    \caption{Edge-state preparation in the AFVH phase. The population of the four left edge modes $L_{3,4,5,6}$, as labeled in Fig.~\ref{fig1}(f), is considered.
    Overlaps of the initial state with Floquet eigenstates are shown with the corresponding wave-packet propagation at the boundary.
    In (a) and (b) [(e) and (f)], the initial state is prepared in the internal spin state $\ket{\eta}=\frac{1}{\sqrt{2}}(1,1)^{\sf T}$ with an initial momentum $(q_x,q_y)=(0.3,0)$ [$(0.3,\pi)$].
    In (c) and (d) [(g) and (h)], the initial state is prepared in $\ket{\eta}=\frac{1}{\sqrt{2}}(1,-1)^{\sf T}$ [$\frac{1}{\sqrt{2}}(1,1)^{\sf T}$] with an initial momentum $(q_x,q_y)=(2.7,0)$ [$(2.85,\pi)$].
    Here, we take $t_{\rm so}=0.5t_0$, $m_{z}=3t_0$, $V_0=3t_0$, $\omega=4t_0$, and $(\delta_x,\delta_y)=(3,1)$.
   }\label{fig5}
\end{figure*}

To illustrate our analysis, we present in Fig.~\ref{fig4}(a)-(d) the analytical solutions for the density distribution $|\phi_L(y)|^2$ of the left edge states at $k_x=0,\pi$ based on Eq.~\eqref{phi_y1}, alongside the numerical results in Fig.~\ref{fig4}(e)-(h), with the phase distribution $\theta(y)$ displayed in the insets. It is evident that the analytical and numerical solutions are in good agreement. Notably, the 0-gap edge-state wave functions at $k_x=0,\pi$ exhibit equal phases at the two leftmost sites $y=1$ and $y=2$, enabling the preparation of a 0-gap edge state without requiring a kick in the $y$-direction. In contrast, the $\pi$-gap edge-state wave functions at $k_x=0,\pi$ exhibit a $\pi$-phase jump at the two leftmost sites, indicating that the wave function is distributed nearly antisymmetrically across these two sites; therefore, an initial kick of $q_y=\pi$ is necessary to prepare an edge state in the $\pi$ gap.

We also note that visible discrepancies exist between analytical and numerical results for the phase distribution $\theta(y)$ in Fig.~\ref{fig4}(a), (d), (e), (h).
The observed deviations arise from two principal sources:
(i) The analytical expressions for edge states in our framework are derived under specific approximations. In particular, the effective Hamiltonian in Eq.~\eqref{Heff_S} that governs driving-induced edge modes was constructed using Floquet perturbation theory~\cite{Zhang2020}.
(ii) The boundary-localized nature of the edge-state density profile introduces enhanced numerical uncertainties at lattice sites far from the boundary ($y\gg1$). 
Nevertheless, both approaches exhibit strong consistency within the near-boundary region ($y\leq2$), which constitutes the primary region of interest in our study.

\begin{figure*}
    \includegraphics[width=\textwidth]{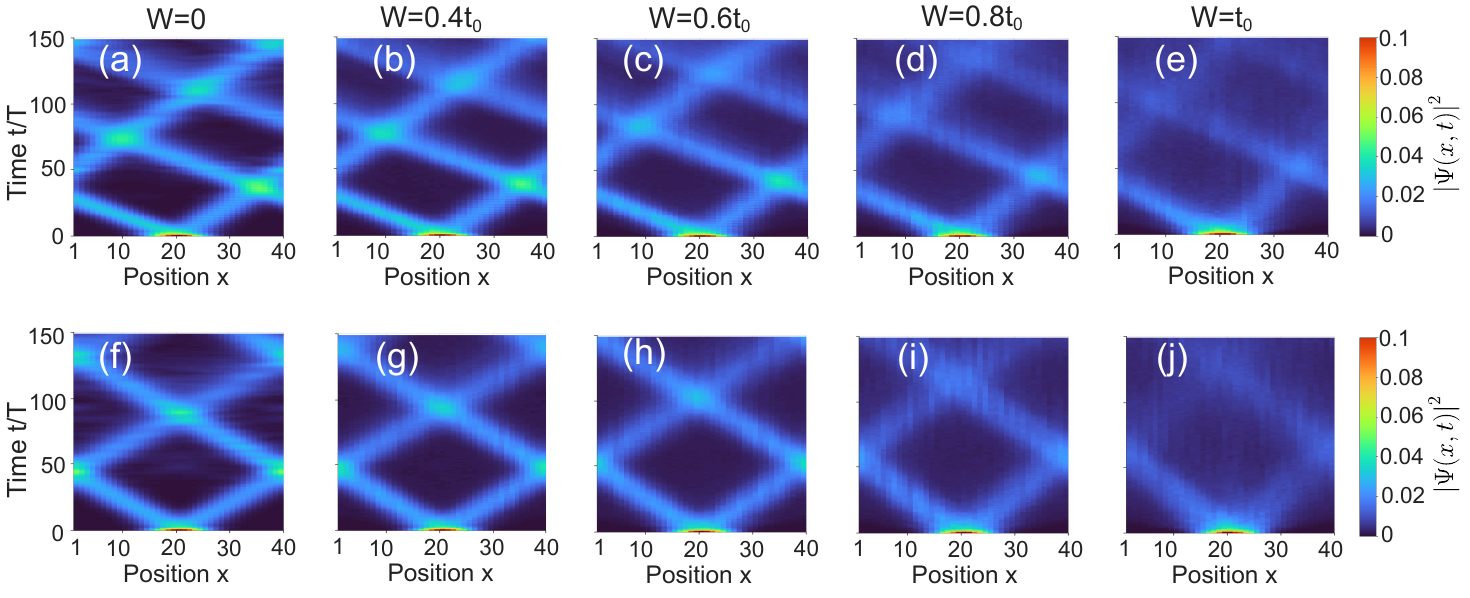}
    \caption{Wave-packet dynamics of counterpropagating edge modes in the 0 gap [(a)-(e)] or $\pi$ gap [(f)-(j)] under the on-site disorder $ V_{\rm rand}(\vb{r})$, with the disorder strength $W$ ranging from $0$ to $t_0$.  Other parameters in (a)-(e) [(f)-(j)] are the same as those in Figs.~\ref{fig5}(b) and \ref{fig5}(d) [Figs.~\ref{fig5}(f) and \ref{fig5}(h)]. Each data point is averaged over 20 disorder configurations.}
    \label{fig6}
\end{figure*}

\begin{figure*}
    \includegraphics[width=\textwidth]{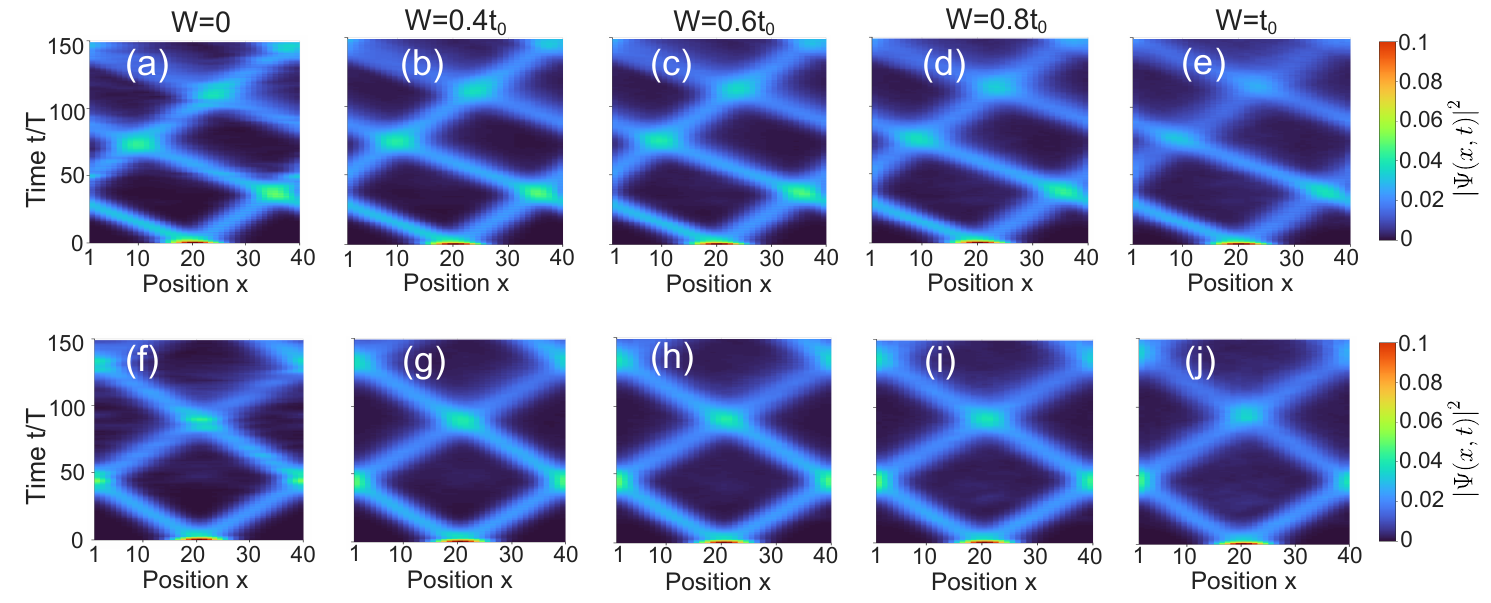}
    \caption{Wave-packet dynamics of counterpropagating edge modes in the 0 gap [(a)-(e)] or $\pi$ gap [(f)-(j)] under the long-range disorder $ V_{\rm sm}(\vb r)$, with the disorder strength $W$ ranging from $0$ to $t_0$.  Other parameters in (a)-(e) [(f)-(j)] are the same as those in Figs.~\ref{fig5}(b) and \ref{fig5}(d) [Figs.~\ref{fig5}(f) and \ref{fig5}(h)]. Each data point is averaged over 20 disorder configurations.}
    \label{fig7}
\end{figure*}

We investigate the overlaps and the chiral motion of the wave packets to demonstrate edge-state preparation in the same way as in the ${\cal C}=2$ phase. Figures~\ref{fig5}(a)-(d) illustrate the initial states prepared in the 0 gap, with Figs.~\ref{fig5}(a) and \ref{fig5}(c) showing significant overlaps with the zero quasienergy modes. Figures~\ref{fig5}(b) and \ref{fig5}(d) highlight the dynamics of the wave packets, which propagate with opposite chirality, confirming that by adjusting different initial momenta $q_x$, the states are effectively prepared to occupy two distinct edge modes $L_{3,5}$ located at $k_x=0$ and $\pi$, respectively. Interestingly, Figs.~\ref{fig5}(e)-(h) depict the initial states prepared in the $\pi$ gap, which require an initial kick of $q_y=\pi$. Figures~\ref{fig5}(e) and \ref{fig5}(g) demonstrate that these initial states exhibit strong overlaps with the $\pi$ quasienergy modes, while Figs.~\ref{fig5}(f) and \ref{fig5}(h) illustrate the wave packets' dynamics, once again showing propagation with opposite chirality. This behavior arises from the two different edge modes $L_{4,6}$ being populated through distinct initial momenta $q_x$.

Note that the selective population of edge modes within distinct quasienergy gaps relies on the symmetric or antisymmetric distribution of the edge-state wave function at the two leftmost sites, which imposes a constraint on the wave-packet width $\delta_y$ of the initial prepared state; specifically, $\delta_y$ must not be too small. In practice, 
$\delta_y$ considerably affects edge-state preparation, with an optimal range found to be between 0.7 and 1 (see Appendix~\ref{App3} for details).

After demonstrating the effect of the initial momentum $q_y$, we now return to the preparation of edge states in the ${\cal C}=2$ phase.
The edge mode $L_2$ in Fig.~\ref{fig1}(d) corresponds to the driving-induced $\pi$ BIS of order $n=1$, for which the effective Hamiltonian is defined with
$\widetilde{m}_z=m_z-\omega/2=-t_0$, within the range $-2t_0\leq \widetilde{m}_z<0$. 
According to Eq.~\eqref{thetay_pi}, populating the mode $L_2$ does not require a kick in the $y$-direction, consistent with the numerical results shown in Figs.~\ref{fig3}(c) and~\ref{fig3}(d). However,  it is worth noting that for higher driving frequencies, e.g. $\omega>10t_0$ for $m_z=3t_0$, an initial kick of $q_y=\pi$ may also be necessary in the preparation of the mode $L_2$.

\section{Counterpropagating edge-state transport}\label{Sec4}

In this section, we investigate the counterpropagating motion of the oppositely chiral edge states in the AFVH phase,
and demonstrate their robustness against long-range disorder. 
To achieve it, we employ the density matrix to describe the time evolution of the wave packet. The initial density matrix reads
\begin{equation}
    \rho(0)=p_1\ket{\Psi_1}\bra{\Psi_1}+p_2\ket{\Psi_2}\bra{\Psi_2}
\end{equation}
where $|\Psi_{1,2}\rangle$ are two states taking the form of Eq.~\eqref{inista} 
with different momentum $(q_x,q_y)$ and/or internal spin state $\ket{\eta}$,
and $p_{1,2}$ are, respectively, the probabilities of the atoms occupying these two states.
In our calculations, $p_1=p_2=0.5$, and the parameters of $|\Psi_{1,2}\rangle$ are chosen to align with those in Fig.~\ref{fig5},
ensuring that the initial state can simultaneously populate two counterpropagating edge modes within either the 0 gap or $\pi$ gap.
The time evolution at stroboscopic time $t=nT$ ($n\in\mathbb{Z}$) is given by
\begin{equation}
    \rho(t=nT) = [U(T)]^n\rho(0)[U^{\dagger}(T)]^n
\end{equation}
where the stroboscopic time-evolution operator $U(T)$ is defined in Eq.~\eqref{UT_def}.
Then, the time-dependent density distribution at the boundary can be obtained by
\begin{equation}
    \abs{\Psi(x,t)}^2=\sum_{\sigma=\uparrow,\downarrow}~\sum_{y_{\rm edge}}\mel{x,y,\sigma}{\rho(t)}{x,y,\sigma},
\end{equation}
where $y_{\rm edge}=1,2,3,4$ in our calculations.

Moreover, we examine the robustness of the edge modes' counterpropagating motion against disorders.
We consider two types of disorder potentials applied to the Zeeman term ($m_z\rightarrow m_z+ V_{\rm disorder}$). 
The first one is random on-site disorder, described by
\begin{equation}
    V_{\rm rand}(\vb{r})=\sum_{j\in \rm site} V_j\delta(\vb r- \vb r_j).
\end{equation}
The second is a long-range potential induced by smooth impurities,
\begin{equation}
    V_{\rm sm}(\vb r) = \sum_{l}^{N_{\rm imp}} \frac{V_l}{\sqrt{(\vb r - \vb r_l)^2+d^2}},
\end{equation}
where $N_{\rm imp}$ is  the number of randomly distributed impurities, $\vb r_l$ denotes the location of $l$th impurity, and $d$ controls the potential range.
Here, we restrict $V_{\rm rand},V_{\rm sm}\in [-W,W]$ with $W$ denoting the disorder strength.

In Fig.~\ref{fig6}, we present the wave-packet dynamics of the counterpropagating edge modes both within the $0$ gap and $\pi$ gap under random disorder  $V_{\rm rand}(\vb{r})$ with the strength $W$ ranging from $0$ to $t_0$. As the on-site disorder is applied and gradually intensified, we observe a progressive decline in the robustness of the edge transport, indicating that the edge states with opposite chirality are scattering into the bulk.
In contrast, Fig.~\ref{fig7} shows that the counterpropagating transport of these edge modes exhibits remarkable robustness against the long-range disorder $V_{\rm sm}(\vb r)$, even as the disorder strength $W$ varies from $0$ to $t_0$. This confirms the prediction made in Ref.~\cite{Zhang2022} that the counterpropagating edge states in the AFVH phase are immune to long-range disorder.

\begin{figure}
    \centering
    \includegraphics[width=0.49\textwidth]{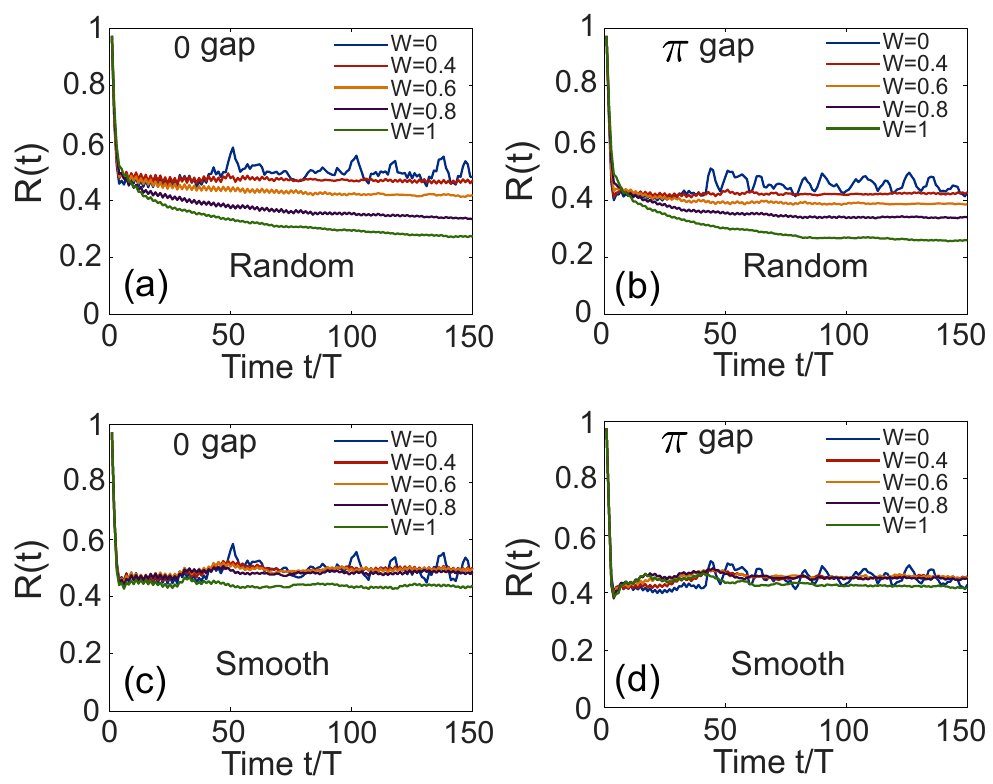}
    \caption{Remnant density $R(t)$ of the wave packet at the boundary as a function of time $t$ under the on-site random disorder $V_{\rm rand}(\vb{r})$ [(a) and (b)] 
    or long-range disorder $V_{\rm sm}(\vb r)$ [(c) and (d)].
    The results are extracted from the data in Fig.~\ref{fig6} and Fig.~\ref{fig7}.}
    \label{fig8}
\end{figure}

To provide a more detailed analysis, we calculate the remnant density of the wave packet at the boundary and present the results in the Fig.~\ref{fig8}. We define $R(t)$ to represent the time-evolving density remaining at the boundary:
\begin{equation}
    R(t)=\sum_{x}\abs{\Psi(x,t)}^2.
\end{equation}
In Figs.~\ref{fig8}(a) and \ref{fig8}(b), we observe that when the on-site random disorder is applied and the disorder strength is increased, the remnant edge-state density decreases. 
Conversely, in Figs.~\ref{fig8}(c) and \ref{fig8}(d),  the remnant edge-state density remains nearly constant as the long-range disorder is applied and intensified.
These results clearly demonstrate that the counterpropagating edge modes with opposite chirality within the $0$ or $\pi$ gap are robust against long-range disorder, while they are significantly affected by on-site random disorder. 

\begin{figure}
    \centering
    \includegraphics[width=0.35\textwidth]{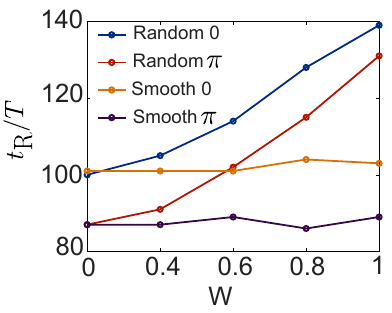}
    \caption{Return time $t_{\rm R}$ of the wave packet prepared in the 0  or $\pi$ gap in the presence of different disorder potentials.
    The result is extracted from the data in Fig.~\ref{fig6} and Fig.~\ref{fig7}.}
    \label{fig9}
\end{figure}

In addition to the analysis presented above, we offer a complementary perspective to investigate the impact of disorder on the edge-state transport.
Specifically, we define a return time $t_{\rm R}$ as the time required for a wave packet to complete a closed loop trajectory 
along the boundary in the periodic $x$-direction.
This parameter inversely quantifies chiral edge-state propagation velocity: A larger $t_R$ corresponds to slower wave-packet motion.
In Fig.~\ref{fig9}, we show the return time for right-propagating wave packets prepared within the 0 and $\pi$ gap under different disorder potentials.
The results reveal that $t_{\rm R}$ increases significantly with stronger on-site random disorder, but remains nearly unchanged in the presence of smooth disorder.
This stark contrast indicates that localized disorder strongly suppresses edge-state transport, while smooth disorder maintains propagation efficiency.
These findings directly support our earlier conclusion that counterpropagating edge modes remain robust against smooth disorder.

\section{Discussion and conclusions}\label{Sec5}

In this work, we provide a detailed investigation of how to prepare and observe the anomalous counterpropagating edge states with
ultracold atoms trapped in a 2D shaken optical Raman lattice.
Starting with an initial Gaussian state, we show that both the internal spin state and the wave-packet's momentum 
must be carefully prepared to achieve effective population of a target edge state.
In particular, in preparing counterpropagating edge states, 
the population within the 0 gap or $\pi$ gap can be effectively 
controlled by the application of an initial kick of $\pi$ 
in the direction perpendicular to the boundary. 
By examining the wave-packet dynamics, 
we demonstrate that the counterpropagating edge-state transport is immune to long-range disorder.
Furthermore, we find that counterpropagating edge states exhibit robustness
even under finite interaction strengths. An investigation of this interaction-dependent stability is provided in Appendix~\ref{App4}.

The preparation and detection scheme discussed here is readily implementable in current experimental setups. 
First, a tightly focused optical tweezer can precisely position the atomic cloud at the system edge~\cite{Braun2024}. 
Second, the internal spin state of the atoms can be controlled via an appropriately tuned Raman-coupling pulse, 
while an initial momentum kick can be introduced by shifting the tight trap either parallel or perpendicular to the system edge prior to release. 
Third, the counterpropagating motion along the edge, observed after releasing the atoms from the trap, can be probed and imaged at a microscopic scale~\cite{Braun2024,Gross2021}, 
providing high-resolution insights into the edge-state transport. 
Additionally, disorder potentials can be introduced and finely controlled by projecting laser speckles~\cite{Lye2005} or by using programmable digital micromirror devices~\cite{Choi2016} or spatial light modulators~\cite{White2020}, which offer customizable and tunable disorder landscapes essential for studying disorder effects.

It is worth noting that the AFVH phase exhibits a dependence on edge geometry: for instance, the counterpropagating edge modes within the 0 gap can vanish on zigzag edges when $V_0>0.6\omega$~\cite{Zhang2022}. In this study, we focus on edge-state transport along a straight edge for simplicity. However, the methods for preparing and observing edge modes discussed here can be readily adapted to other edge geometries. Experimentally, the shape and orientation of the edge can be controlled using a programmable repulsive optical potential~\cite{Braun2024}, allowing for versatile exploration of geometry-dependent effects in the AFVH phase.

In summary, we have presented a detailed study on the preparation and observation of anomalous counterpropagating edge states 
with ultracold atoms in a periodically driven optical Raman lattice, 
establishing a framework for future experimental investigations into the AFVH phase and its unique topological characteristics.

\section*{Acknowledgements}

H. H. thanks Xiao-Dong Lin and Bei-Bei Wang for helpful discussions.
This work was supported by the National Natural Science Foundation of China (Grants No.~12204187),  
the Innovation Program for Quantum Science and Technology (Grant No.~2021ZD0302000), 
and the startup grant of Huazhong University of Science and Technology (Grant No.~3034012114).

\begin{appendix}

\section{Band-inversion surfaces and topological classification theory}\label{App1}

In this Appendix, we briefly review the main results of the BIS-based classification theory. 
More details can be found in Refs.~\cite{Zhang2018,Zhang2019a,Zhang2019b,Zhang2020}.
For a static system $H_s(\bold{k})$, the component $h_z(\bold{k})$ can be chosen to describe the dispersion.
The BIS refers to the momenta where band inversion occurs, i.e.,
\begin{align}~\label{BIS_def_S}
{\rm BIS}\equiv\{\bold{k}| h_z(\bold{k})=0\}. 
\end{align} 
The remaining components $h_{x,y}(\bold{k})$ composes a spin-orbit (SO) field 
$\bold{h}_{\rm so}(\bold{k})=(h_y,h_x)$, which opens a gap at BISs and leads to nontrivial band topology.
According to the bulk-surface duality~\cite{Zhang2018}, the Chern number that characterizes the bulk topology has a one-to-one correspondence to the 1D winding number defined on all BISs, i.e., 
\begin{align}
\mathcal{C}=\sum_{i}\nu_{i},
\end{align}
where $\nu_i$ counts the winding of the SO field along the $i$th BIS:
\begin{equation}
\nu_{i}=\frac{1}{2\pi}\int_{{\rm BIS}_j} d{k_\parallel}\left(\hat{h}_y\partial_{{k}_\parallel} \hat{h}_x - \hat{h}_x\partial_{{k}_\parallel}\hat{h}_y\right),
\end{equation}
with $k_\parallel$ denoting the momentum along the BIS and $\hat{h}_{\alpha}={h}_{\alpha}/\sqrt{h_x^2+h_y^2}$ ($\alpha=x,y$).

The above characterization can be generalized to Floquet topological phases~\cite{Zhang2020}. 
For the 2D Floquet Hamiltonian taking the form of Eq.~\eqref{HF_general}, one can introduce the BISs of Floquet bands as
\begin{align}~\label{FBIS_def_S}
{\rm BIS}\equiv\{\bold{k}| h_{F, z}(\bold{k})=0\},
\end{align} 
and accordingly define the Floquet SO field $\bold{h}_{F, {\rm so}}(\bold{k})\equiv(h_{F, y}, h_{F, x})$.
Due to the periodicity of quasienergy,
Floquet band crossings can appear in both the 0 and $\pi$ gaps, and the BISs are determined by~\cite{Zhang2020} 
\begin{align}
h_z(k)=n\omega/2,\quad  n=0,1,2,\cdots,
\end{align}
where the one of even (odd) order $n$ is the one associated with the 0 gap ($\pi$ gap), named as $0$ BIS ($\pi$ BIS).
The topology of the Floquet band below the $0$ gap is contributed by all $0$ and $\pi$ BISs but with opposite signs~\cite{Zhang2020}:
\begin{equation}~\label{W_Floquet_S}
\mathcal{C}=\mathcal{W}_0-\mathcal{W}_{\pi}, \quad \mathcal{W}_q = \sum_{i}\nu_{i}^{(q)},
\end{equation}
where ${\cal W}_{0}$ (${\cal W}_{\pi}$) characterizes the number of boundary modes inside the 0 gap ($\pi$ gap)~\cite{Rudner2013}, and
$\nu_{i}^{(q)}$ represents the topological invariant defined on the $i$th $q$ BIS ($q = 0, \pi$): 
\begin{align}
\nu_{i}^{(q)} =\frac{1}{2\pi}\int_{q\,{\rm BIS}_i} d{k_\parallel}\left(\hat{h}_{F,y}\partial_{{k}_\parallel} \hat{h}_{F,x}- \hat{h}_{F,x}\partial_{{k}_\parallel}\hat{h}_{F,y}\right).
\end{align}

In addition to characterizing the bulk topology, the local topology on each BIS 
is uniquely linked to the emergence of gapless boundary modes~\cite{Zhang2022}. 
It has been shown that a Floquet system can be decomposed into multiple static subsystems, 
each periodic in quasienergy and composed of bands inverted on a BIS~\cite{WangBB2024}.
For the 2D driven system described by Eq.~\eqref{FHam2D},
the subsystem that corresponds to the BIS of order $n$
can be described by the following effective static Hamiltonian~\cite{Zhang2022}
\begin{equation}~\label{Heff_S}
H_{\rm eff}^{(n)}=\left(h_z-\frac{n\omega}{2}\right)\sigma_z+(-1)^nJ_n\left(\frac{4V_0}{\omega}\right)(h_x\sigma_x+h_y\sigma_y),
\end{equation}
where $J_n(z)$ denotes the Bessel function of the first kind.
This effective Hamiltonian takes the same form as the undriven Hamiltonian $H_s$ under parameter substitutions
\begin{align}\label{meff_replace_S}
m_z\to \widetilde{m}_z=m_z-\frac{n\omega}{2},
\end{align}
and
\begin{align}\label{tso_replace_S}
t_{\rm so}\to\tilde{t}_{\rm so}=(-1)^nJ_n\left(\frac{4V_0}{\omega}\right)t_{\rm so},
\end{align}
thus supporting a gapless edge mode when $0<|\widetilde{m}_z|<4t_0$.
This result reveals the one-to-one correspondence between the topology on BISs and gapless edge modes at the boundary.
In this work, we employ the effective subsystem Hamiltonian \eqref{Heff_S} to analyze the driving-induced edge modes.

\section{Analytical expressions for chiral edge modes}~\label{App2}

In this Appendix, we drive analytical expressions for the gapless edge modes in both static and periodically driven systems.
We first consider the static Hamiltonian $H_{\rm s}({\bf k})$ and impose open boundary conditions in the $y$-direction.
The Hamiltonian can be rewritten as
\begin{align}
    H(k_x)=\sum_{y}{\cal E}(k_x) c^{\dagger}_{y}c_{y}+\sum_y {\cal T}_y c^{\dagger}_{y+1}c_{y}+{\rm h.c.}
\end{align}
where
\begin{align}
    &{\cal E}(k_x)=(m_{z}-2t_0\cos k_x)\sigma_{z}+2t_{\rm so}\sin k_x\sigma_{x}, \nonumber\\
    &{\cal T}_y=it_{\rm so}\sigma_y-t_0\sigma_{z}\nonumber.
\end{align}
We assume that the edge-state wave functions take the form
\begin{align}\label{edge_wavefun_S}
|\phi_{L/R}(k_x,y)\rangle=\frac{u_{k_x}(x)}{\mathcal{N}}\lambda^{y}|\eta_{L/R}\rangle,
\end{align}
where $\lambda$ is a complex number, ${\cal N}$ is the normalization factor, $u_{k_x}(x)$ denotes the Bloch wave function along the $x$-direction, 
and $|\eta_{L/R}\rangle$ are two-component spinors. 
The eigenvalue equation $H(k_x)\ket{\phi}=E\ket{\phi}$ leads to
\begin{equation}\label{2dstegeq_S}
    ({\cal T}_y^{\dagger}\lambda^{-1}+{\cal E}(k_x)+{\cal T}_y\lambda)\ket{\eta}=E\ket{\eta},
\end{equation}
or, in a matrix form,
\begin{widetext}
\begin{align}\label{2dstegeq_mx_S}
    &\begin{pmatrix}
        -2 t_0 \cos k_x+m_z-\left(\lambda +\lambda^{-1}\right) t_0 & 2 t_{\text{so}} \sin k_x+\left(\lambda-\lambda^{-1}\right) t_{\text{so}}\\
        2 t_{\text{so}} \sin k_x+\left(\lambda^{-1}-\lambda\right) t_{\text{so}} & 2t_0 \cos k_x-m_z+\left(\lambda+\lambda^{-1}\right) t_0
       \end{pmatrix}
       \begin{pmatrix}
        \alpha \\\beta 
       \end{pmatrix}=E
       \begin{pmatrix}
        \alpha \\\beta 
       \end{pmatrix},
\end{align}
which gives the eigenvalues
\begin{equation}\label{energy_kx}
    E(k_x,\lambda)=\pm\sqrt{\left[(m_z-2t_0\cos k_x)-t_0(\lambda+\lambda^{-1})\right]^2+4t_{\rm so}^2\sin^2k_x-t_{\rm so}^2(\lambda-\lambda^{-1})^2}.
\end{equation}
Since $E(k_x,\lambda)=\pm2t_{\rm so}\sin k_x$~\cite{Mong2011},
we have $\left[(m_z-2t_0\cos k_x)-t_0(\lambda+\lambda^{-1})\right]^2-t_{\rm so}^2(\lambda-\lambda^{-1})^2=0$,
which yields four solutions:
\begin{align}\label{lambda_solutions_S}
\begin{split}
\lambda_{1\pm}&=\frac{m_z-2t_0\cos k_x\pm\sqrt{\left(m_z-2t_0 \cos k_x\right)^2-4 \left(t_0^2-t_{\text{so}}^2\right)}}{2(t_0+t_{\text{so}})},\,{\rm with}\,
\ket{\eta_1}=\frac{1}{\sqrt{2}}
\begin{pmatrix}
       1 \\ -1
 \end{pmatrix},\, E=-2t_{\rm so}\sin k_x;\\
\lambda_{2\pm}&=\frac{m_z-2t_0\cos k_x\pm\sqrt{\left(m_z-2t_0 \cos k_x\right)^2-4 \left(t_0^2-t_{\text{so}}^2\right)}}{2(t_0-t_{\text{so}})}=1/\lambda_{1\mp},\,{\rm with}\,
\ket{\eta_2}=\frac{1}{\sqrt{2}}
\begin{pmatrix}
       1 \\ 1
 \end{pmatrix},\, E=2t_{\rm so}\sin k_x.
\end{split}
\end{align}
\end{widetext}

\begin{figure}
    \centering
    \includegraphics[width=0.4\textwidth]{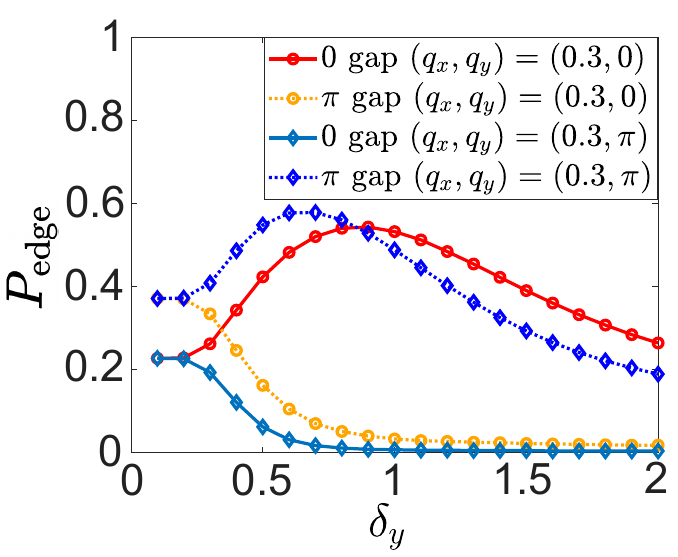}
    \caption{Probability of populating two edge modes $L_{3,4}$ with different wave-packet width $\delta_y$.
    The solid (dashed) lines represent the results of an initial state with momentum $(q_x,q_y)=(0.3,0)$ or $(0.3,\pi)$ populating the 0-gap edge mode $L_3$ (the $\pi$-gap edge mode $L_4$). Other parameters are the same as those in Figs.~\ref{fig5}(a) and \ref{fig5}(e).}
    \label{fig10}
\end{figure}

Since the wavefunctions in Eq.~\eqref{edge_wavefun_S} decay (or increase) with $y$ when $|\lambda|<1$ (or $|\lambda|>1$ ), 
the edge mode localized at the left (right) boundary corresponds to solutions with $|\lambda|<1$ ($|\lambda|>1$). 
Therefore, the left edge mode takes the form
\begin{align}\label{left_wavefun_S}
\ket{\phi_L(k_x,y)}=\frac{u_{k_x}(x)}{\mathcal{N}_L}\left(\lambda_{L+}^{y}-\lambda_{L-}^{y}\right)\ket{\eta_L},
\end{align}
where $\lambda_{L\pm}$, with $|\lambda_{L,\pm}|<1$, can be $\lambda_{1\pm}$ or $\lambda_{2\pm}$, 
which depends on the parameters and also determines the internal state $\ket{\eta_L}$.
For example, when $m_z=3t_0$ ($m_z=-t_0$) and $t_{\rm so}=0.5t_0$, there exists a left edge mode around $k_x=0$ ($k_x=\pi$), 
which is described by Eq.~\eqref{left_wavefun_S} with
$\lambda_{L\pm}=\lambda_{1\pm}$, $\ket{\eta_L}=\ket{\eta_1}$, and has negative (positive) chirality with $E=-2t_{\rm so}\sin k_x$.

We then consider the Floquet Hamiltonian \eqref{FHam2D}. According to the BIS-boundary correspondence established in Ref.~\cite{Zhang2022}, 
each 0 or $\pi$ BIS with nontrivial topology corresponds to a chiral edge mode within the corresponding gap.
Such a correspondence is reflected in the effective Hamiltonian \eqref{Heff_S}. 
Consequently, the driving-induced left edge modes also take the form of Eq.~\eqref{left_wavefun_S} with
$\lambda_{L\pm}$ being either $\lambda_{1\pm}$ or $\lambda_{2\pm}$ after the substitutions specified in Eqs.~\eqref{meff_replace_S} and \eqref{tso_replace_S}.

Here, we take the four left edge modes $L_{3,4,5,6}$ in Fig.~\ref{fig1}(f) as illustrative examples.
The existence of the edge mode $L_3$ is ensured by the nonzero topological invariant defined on the static $0$ BIS (the order $n=0$). 
The effective subsystem Hamiltonian that manifests this BIS-boundary correspondence is then defined with $\widetilde{m}_z=m_z=3t_0$ and $\tilde{t}_{\rm so}=J_0\left(3\right)t_{\rm so}\approx-0.26t_{\rm so}$. As a result, the edge mode $L_3$ is described by Eq.~\eqref{left_wavefun_S} 
with $\lambda_{L\pm}=\lambda_{2\pm}$ and $\ket{\eta_L}=\ket{\eta_2}$.
The left edge mode $L_4$ corresponds to the driving-induced $\pi$ BIS of order $n=1$, for which $\widetilde{m}_z=m_z-\omega/2=t_0$ 
and $\tilde{t}_{\rm so}=-J_1\left(3\right)t_{\rm so}\approx-0.34t_{\rm so}$. This results in the left edge mode $L_4$ being located around $k_x=0$ within the $\pi$ gap and taking the form of Eq.~\eqref{left_wavefun_S} also with $\lambda_{L\pm}=\lambda_{2\pm}$ and $\ket{\eta_L}=\ket{\eta_2}$.
The left edge mode $L_5$ corresponds to the $0$ BIS of order $n=2$; the parameters $\widetilde{m}_z=m_z-\omega=-t_0$ 
and $\tilde{t}_{\rm so}=J_2\left(3\right)t_{\rm so}\approx0.49t_{\rm so}$ determine that $L_5$ is located around $k_x=\pi$ within the $0$ gap and exhibits a distribution characterized by $\lambda_{L\pm}=\lambda_{1\pm}$ and $\ket{\eta_L}=\ket{\eta_1}$.
The left edge mode $L_6$ corresponds to the $\pi$ BIS of order $n=3$; the parameters
$\widetilde{m}_z=m_z-3\omega/2=-3t_0$ and $\tilde{t}_{\rm so}=-J_3\left(3\right)t_{\rm so}\approx-0.31t_{\rm so}$ lead to $L_6$
distributed around $k_x=\pi$ within the $\pi$ gap, with $\lambda_{L\pm}=\lambda_{2\pm}$ and $\ket{\eta_L}=\ket{\eta_2}$.
The above analytical predictions agree with the numerical results.

Moreover, it is noteworthy that when the driving strength is relatively weak, $V_0<0.6\omega$, ensuring $J_n\left(\frac{4V_0}{\omega}\right)>0$ for all $n\leq0$,
the sign of the parameter $\tilde{t}_{\rm so}$ depends solely on the order $n$ [cf. Eq.~\eqref{tso_replace_S}].
When $n$ is even, we have $\tilde{t}_{\rm so}>0$ and the solutions with $|\lambda|<1$ can only be $\lambda=\lambda_{1\pm}$ with the spin state $\ket{\eta_1}$.
In contrast, when $n$ is odd, we have $\tilde{t}_{\rm so}<0$ and the solutions with $|\lambda|<1$ are $\lambda=\lambda_{2\pm}$ with the spin state $\ket{\eta_2}$.
This leads to the general conclusion presented in Eq.~\eqref{etaL}.

\section{The dependence of edge-state preparation on the width $\delta_y$}~\label{App3}

Throughout the main text, the wave-packet width is fixed at $\delta_y=1$.
In this Appendix, we investigate how varying the width $\delta_y$ affects the preparation of edge states.
To this end, we calculate the total probability of the prepared initial state populating the target edge mode $L_i$ by
\begin{align}
P_{\rm edge}=\sum_{n\in L_i}\sum_{x,y}|\braket{\phi_n}{\Psi_0}|^2,
\end{align}
where $\ket{\phi_n}$ denote the eigenstates of the edge mode $L_i$ and $\ket{\Psi_0}$ is the initial state taking the form of Eq.~\eqref{inista}.

We revisit the preparation of the edge modes $L_{3,4}$ shown in Figs.~\ref{fig5}(a) and \ref{fig5}(e),
and present the calculated probability $P_{\rm edge}$ in Fig.~\ref{fig10}, where the width $\delta_y$ is varied from 0.1 to 2.
We find that (i) the probability of populating the target edge modes $L_{3,4}$ depends on the width $\delta_y$,
with an optimal range between 0.7 and 1; and (ii) the momentum $q_y$ can effectively
control the population of the edge mode within the 0 or $\pi$ gap, requiring the width $\delta_y>0.6$. 
This second result is consistent with our analysis, in which it is the wave-function distribution 
at the two leftmost sites that results in the selective population via an initial kick $q_y$.

\section{Stability of Floquet edge states under interactions}~\label{App4}

\begin{figure}
    \centering
    \includegraphics[width=0.49\textwidth]{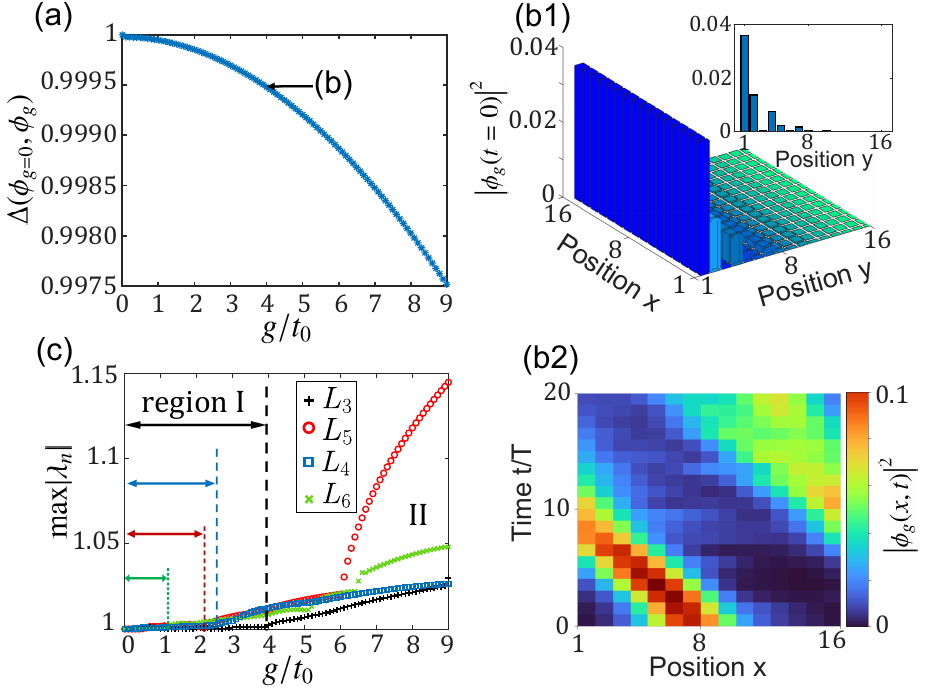}
    \caption{Interaction-dependent stability of anomalous counterpropagating edge states in the AFVH phase.
    (a) Fidelity $\Delta(\phi_{g = 0}, \phi_g)$ versus interaction strength $g$, where $|\phi_{g = 0}\rangle$ represents an edge state of the $L_5$ mode.
    (b) Edge-state characterization of $|\phi_{g=4t_0}\rangle$ (indicated by a black arrow): Boundary-confined spatial density profile in (b1) and chiral transport probed through the Gaussian wave-packet dynamics in (b2).
    (c) Maximum eigenvalues $|\lambda_n|$ of operator $G(T)$ as functions of $g$ for all counterpropagating edge modes $L_{3,4,5,6}$. 
     Two distinct regions can be identified: Long-lifetime region I ($\max|\lambda_n|\approx1$) and short-lifetime region II ($\max|\lambda_n|>1$).}
    \label{fig11}
\end{figure}

We employ the Gross-Pitaevskii equation framework described in Ref.~\cite{Mochizuki2021} to investigate the stability of Floquet edge states under interactions. 
For simplicity, we consider a Bose gas with spin-independent contact interactions.

For an edge eigenstate $|\phi_{g=0}\rangle$ of the non-interacting Floquet Hamiltonian $H_F$,
the corresponding stationary edge states $|\phi_{g}\rangle$ with different interaction strengths $g$ are obtained through the following procedure: 
First, we construct the nonlinear time-evolution operator
\begin{align}
    U_g(t)=\mathcal{T}e^{-\ui\int_{0}^{t}  H_{g}(\tau) d\tau},
\end{align}
where the interacting Hamiltonian reads
\begin{align}
H_{g}(t)=H(t)+\sum_{\sigma=\uparrow,\downarrow}\sum_{{\bf r}}g|\phi_{g}^{\sigma}({\bf r},t)|^{2}|{\bf r}\rangle\langle{\bf r}|,
\end{align}
with $|\phi_{g}(t)\rangle=U_g(t)|\phi_{g}\rangle=\left(\phi_{g}^{\uparrow}(t),\phi_{g}^{\downarrow}(t)\right)^{\sf T}$.
The diagonalization of the one-period evolution operator $U_g(T)$ yields
\begin{equation}
    U_g(T)\ket{\phi_n}=e^{-\ui \varepsilon_n T}\ket{\phi_n}.
\end{equation}
We then calculate the fidelity $\Delta(\phi_{g=0},\phi_n)$ for all $\ket{\phi_n}$:
\begin{equation}\label{fidelity_Delta}
    \Delta(\phi_{g=0},\phi_n)=\frac{|\langle\phi_{g=0}|\phi_n\rangle|}{\sqrt{\langle\phi_{g=0}|\phi_{g=0}\rangle\langle\phi_n|\phi_n\rangle}},
\end{equation}
and identify $|\phi_g\rangle$ as the eigenstate $|\phi_{n=n^\star}\rangle$ with the maximum fidelity.
This process is iteratively repeated using $|\phi_{g-\delta g}\rangle$ (with small $\delta g$) as the initial state for the
first iteration.


According the Floquet theorem, the time-periodic eigenstate satisfies $|\phi_g(t)\rangle\equiv|\phi_{n^\star}(t)\rangle=e^{-\ui\varepsilon_{n^\star} t} |u_{n^\star}(t)\rangle$, 
with $|u_{n^\star}(t)\rangle=|u_{n^\star}(t+T)\rangle$~\cite{Eckardt2017_review}.
The stability of the Floquet stationary state $|\phi_g(t)\rangle$ can be characterized through the eigenvalues $\{\lambda_n\}$ of the non-unitary Floquet operator
\begin{align}
  G(T)=\mathcal{T}e^{-\ui\int_{0}^{T} K(t) dt},
 \end{align}
where the non-Hermitian operator $K(t)$ reads
\begin{align}
    K(t)&=\begin{pmatrix}
    Q(t)& S(t)\\
    -S^*(t)& -Q(t)
    \end{pmatrix},
\end{align}
with
\begin{align}
\begin{split}
    Q(t)&=H(t)+\sum_{\sigma=\uparrow,\downarrow}\sum_{{\bf r}}\left[2g\left|u_{n^\star}^\sigma({\bf r},t)\right|^{2}-\varepsilon_{n^\star}\right]|{\bf r}\rangle\langle{\bf r}|,\\
    S(t)&=\sum_{\sigma=\uparrow,\downarrow}\sum_{{\bf r}}g\left[u_{n^\star}^{\sigma}({\bf r},t)\right]^2|{\bf r}\rangle\langle{\bf r}|.\\
   \end{split}
\end{align}
The state $|\phi_g(t)\rangle$ remains dynamically stable when all $\lambda_n$ satisfy $|\lambda_n|<1$~\cite{Mochizuki2021}.

We adopt the described protocol to investigate the interaction-dependent stability of anomalous counterpropagating edge states.
Through fidelity analysis [Eq.~\eqref{fidelity_Delta}], we first identify stationary edge states $|\phi_g\rangle$   
that inherit topological characteristics from their non-interacting counterparts $|\phi_{g=0}\rangle$, as exemplified in Fig.~\ref{fig11}(a) for the $L_5$ mode.
To confirm the edge-state nature, we perform two complementary diagnostics: 
(i) The boundary-confined density profile of $|\phi_{g=4t_0}\rangle$ is shown in Fig.~\ref{fig11}(b1), 
and (ii) its chiral propagation is revealed through real-space wave-packet dynamics in Fig.~\ref{fig11}(b2).
Finally, we ewe assess the robustness of these stationary edge states by analyzing the spectral properties of the operator $G(T)$.
Figure~\ref{fig11}(c) presents the maximum eigenvalues $|\lambda_n|$ as functions of interaction strength $g$, 
revealing two distinct regions for all counterpropagating edge modes ($L_{3}$-$L_{6}$): 
Long-lifetime region I with $\max|\lambda_n|\approx1$ and short-lifetime region II with $\max|\lambda_n|>1$.
This bifurcation demonstrates that anomalous counterpropagating edge states in the AFVH phase can maintain robustness within finite interaction ranges.


\end{appendix}


\noindent

\end{document}